\documentclass[10pt,letterpaper,twocolumn]{article}
\usepackage[utf8]{inputenc}
\usepackage[T1]{fontenc}
\usepackage{amsmath,amssymb,amsfonts}
\usepackage{graphicx}
\usepackage{hyperref}
\usepackage{url}
\usepackage{microtype}
\usepackage{multirow}
\usepackage{ragged2e}
\usepackage{verbatim}
\usepackage{multicol}
\usepackage{float}
\usepackage{geometry}
\usepackage{titlesec}
\usepackage{footmisc}
\usepackage{authblk}
\usepackage{enumitem}
\titleformat{\section}
  {\large\bfseries} 
  {\thesection}{1em}{} 

\titleformat{\subsection}
  {\bfseries} 
  {\thesubsection}{1em}{} 

\titleformat{\subsubsection}
  {\itshape} 
  {\thesubsubsection}{1em}{} 

\titlespacing*{\section}
  {0pt}{1.0em}{0.75em} 

\titlespacing*{\subsection}
  {0pt}{0.75em}{0.5em} 

\titlespacing*{\subsubsection}
  {0pt}{0.5em}{0.25em} 

\usepackage{caption}
\title{Plug-and-Play Drag Sail Module for LEO Satellites: Implementation and Early Testing of AirDragMod (ADM)}
\author[1]{Anshuman Shukla\thanks{Corresponding author: \texttt{anshuman1.mitmpl2024@learner.manipal.edu}}}

\author[2]{Pranav Sawant}

\affil[1]{\centering Department of Electrical \& Electronics Engineering, MIT Manipal, Manipal University, Udupi, Karnataka 576104, India\\\vspace{0.5em}
\texttt{anshuman1.mitmpl2024@learner.manipal.edu}}

\affil[2]{\centering Department of Computer Science, The University of Texas at Dallas, Richardson, TX 75080, USA\\\vspace{0.5em}
\texttt{pranav.sawant@utdallas.edu}}

\date{}

\begin{document}
\maketitle

\begin{abstract}
\noindent
\justifying
Space debris has become an increasingly critical subject, with the amount of debris in orbit now exceeding the number of active operational satellites, posing significant risks to the sustainability of space activities. A major constituent of debris is a payload or rocket body discarded in orbit post-mission with no orbital control capacity (OCC) in LEO \cite{ESA_Space_Environment_Report}. The Inter-Agency Space Debris Coordination Committee (IADC) guidelines recommend limiting the post-mission presence of a rocket body or payload in a protected region to 25 years \cite{NASA2020}. Recently, the Federal Communications Commission (FCC) introduced stricter regulations, reducing the allowable post-mission stay to no more than 5 years for satellites in LEO \cite{FCC_Document}. These changes necessitate the integration of deorbiting systems in satellite designs. Adding extra fuel and engines for active thrust-induced deorbiting poses significant challenges. LEO satellites, especially those in large constellations or small CubeSats, i.e., the major composition of LEO\textsubscript{IADC} \cite{NASA2020}, are launched with strict mass and volume limitations. The limited payload capacity of launch vehicles necessitates prioritizing mission-critical components over supplementary systems like deorbiting mechanisms. Therefore, alternative approaches, such as passive deorbiting techniques or international regulatory measures, are often explored to address the issue of space debris in LEO. A well-known cost-effective passive deorbit solution in the field is drag sails owing to their effective method of deorbiting small and medium-sized satellites in LEO. In this paper, we propose a plug-and-play design of a drag sail module, different from the mechanical boom and electrical deployment methods for drag sails currently used, using commercial-off-the-shelf (COTS) components for CubeSats and satellites under 700 kg in mass for LEO\textsubscript{IADC}. 

A conceptual scalable design is proposed, the dimensions of which are derived keeping the mission requirements from two standard sample missions and trajectory analysis results in focus. In the process, a technique to assist in quicker de-orbiting via active control at specific orbital Local Time of Ascending Node (LTAN) was devised. The objective is to develop a ready-to-use, scalable solution for the problem, with COTS components and standard processes for trustworthy acquisition. The deployment mechanism suggested in the paper is based on residual angular momentum inspired by Japan Aerospace Exploration Agency’s (JAXA) IKAROS mission \cite{TSUDA2013183} and was devised along with a standard deployment sequence. A cost analysis was also estimated to infer the breakeven point for such a system. Then, a 3D-printed embodiment of the previously devised deployment system with an inverted stepper motor assembly was incorporated into the prototype to simulate its viability. Following this, tests were conducted over an adjusted version of the standard deployment sequence. The results from these tests were compared and reproduced using a numerical model of the stepper motor. A need was identified to produce a model of the tension varying in the sail extension petals throughout the deployment sequence; the model was produced. Curve fitting techniques were employed to generate a resembling model from the data obtained during test runs. The SIMULINK\texttrademark{} multibody models thus obtained can be used to perform software simulations on the module in a computer-generated environment. Further experimentation is needed, followed by a working prototype, to gauge its working in a real environment. The analysis presented the requirements of a control system, which was found to be of high priority in the development process.
\end{abstract}


\section{Introduction}
\label{sec:introduction}
In recent years, Low Earth Orbit (LEO) satellites have increasingly encountered challenges related to space debris. The inability to track objects smaller than 10 cm in LEO \cite{NASA_OIG_2024} has escalated the risks to critical infrastructure, such as the International Space Station (ISS), necessitating frequent collision avoidance maneuvers. With the new FCC rule, it is only a matter of time before satellite makers are compelled to incorporate deorbiting systems. Heavier satellites can carry extra fuel for deorbiting burns but are constrained by mass. Due to their small size, CubeSats have generally avoided significant encounters with space debris. However, with their increasing deployment, timely deorbiting has become critical. Satellites within constellations both face and pose substantial threats to other satellites. As the population of satellite constellations continues to grow, prompt deorbiting at the end of their operational life has become increasingly essential. Such satellites typically lack active control systems for deorbiting due to limited mass, size, and cost constraints. This reinforces the need for a passive deorbiting system.

Since deorbiting isn’t an integral phase of a mission and doesn’t directly affect its operations, a deorbiting system must be as "invisible" as possible, with low mass, volume, and operational power usage. Although launch costs have significantly fallen, mass still affects satellite power usage during necessary attitude control, thus impacting mission life.

The most implemented deorbiting technique is drag sails. This research paper proposes a drag sail design that is easily scalable, easy to implement, and uses low power. A drag sail deorbiting system uses a thin sail to maximize the drag area of a satellite, increasing the perturbations of aerodynamic drag force.
\setlength{\abovedisplayskip}{5pt}
\setlength{\belowdisplayskip}{0pt}
\begin{equation}
F_d = \frac{1}{2} \rho A C_d v^2
\end{equation}

Equation (1) is the basic equation governing aerodynamic drag on a surface, where \( F_d \) is the drag force, \( \rho \) is the density of the medium, \( A \) is the cross-sectional area perpendicular to the atmospheric flow vector, \( C_d \) is the coefficient of drag (generally taken as 2.2 for LEO), and \( v \) is the orbital velocity of the craft. While Earth's atmosphere thins at higher orbits, it still causes significant degradations in the orbit of LEO satellites over time, necessitating boost burns to stay on course. A drag sail increases the rate of orbital decay, aligning with current Space Debris Mitigation best practices and guidelines \cite{SanchezArriaga}. Furthermore, the change in orbital parameters must exceed the minimum detection capabilities of satellite tracking networks \cite{Lemmens2014TwoLineElementsBasedMD}.

Significant advancements in drag sail deorbiting technology have been achieved over the past decade.\cite{NASA2024DeorbitSystems} Examples include the TechEdSat4 exo-brake launched in 2014, which successfully deorbited with a 0.35 m² drag sail area \cite{TechEdSat4}; TechDemoSat-1 in 2014 with a 6.2 m² drag sail area \cite{TechDemoSat-1}; dragNET\texttrademark{}, which successfully deorbited a Minotaur upper stage in 2016 using a 14 m² drag sail area \cite{OrbitalDebris2019}; CANX-7 (Canadian Advanced Nanosatellite eXperiment-7) in 2017 with a 5 m² drag sail area \cite{Shmuel2012}; and removeDebris in 2018 with a drag sail area of 0.35 m² \cite{AGLIETTI2020310}. There are more missions currently in orbit or planned, including the ADEO-N2 subsystem of 1U (10 cm x 10 cm x 10 cm) size capable of stowage of 5 m² of drag sail by ESA’s Clean Green Space Initiative \cite{ADEO-N2021}.

In this paper, we propose a Plug and Play (PnP) module design standardized using COTS components as a ready-to-use solution for deorbiting. The PnP design philosophy enables quick and reliable installation. This deorbiting module has been proposed in two configurations: one for CubeSats and another for heavier satellites of up to approximately 200 kg in an orbit of up to approximately 900 km for deorbit within 5 years or in an orbit of up to 2000 km for deorbit within 20 years depending on the ballistic coefficient. The design is inspired by JAXA's IKAROS mission, which was the first interplanetary solar sail propulsion mission launched in 2012 and performed a flyby of Venus. Weighing 300 kg, it consisted of a 14 m x 14 m spinning sail and used angular momentum leftover after separation from its rocket stage to deploy four tiny masses, which due to centrifugal force, extended deploying the sail attached to it \cite{Mori2014}. This technique is more efficient than electrically or stored mechanical energy deployed systems, as we will further discuss in this paper.

With these considerations, we present AirDragMod (ADM). To justify the design's capability, a trajectory analysis of simulations run in FreeFlyer\texttrademark{} was performed. Based on the results, the design has been proposed and divided into three phases: i. Conceptual design of both configurations, ii. Sizing of drag sail and module dimensions, and iii. Small-scale deployment system prototype testing. Early tests were on an initial prototype of AirDragMod (ADM), providing valuable data. This prototype featured four degrees of freedom, incorporating the deployment mechanism described previously with appropriate sensors to collect data necessary for refining the mechanism and deriving an analytical model of tension force at the motor drive during deployment. The paper concludes with a comparison and cost analysis, followed by a discussion of the scope for further research.

The primary components of the system include the rotator, aid masses, and sail deployment drives. The rotator, controlled by the deployment mechanism, generates the required rotation in sync with a pre-defined sequence. The aid masses are released first; connected to the free end of the sail, these masses create tension that causes the sail to elongate. The motor units drive the sail, effectively "deploying" it at a constant rate until full extension is achieved.

In the prototype setup, the deployment sequence was standardized to 184 seconds, with 120 seconds allocated for sail deployment. The collected data will be utilized to develop an active control system to manage perturbations during deployment. An accurate dynamic model of the forces at the base of the sail extension (at the motor drives) is crucial. This paper derives an analytical model of tension resulting from centripetal forces acting on the sail petals, using flex sensors in the experimental setup. Additionally, angular velocities were interpolated, using data from IMUs, and vibration data from the system was visualized on a surface plot with the derived model. This analysis helps identify periods of amplified vibration during the deployment sequence.

\section{Requirements Analysis Using Prior Work}
\label{sec:background}

While the ultimate requirements remain consistent due to ADM's scalability, two distinct reference studies have been chosen to address the design needs. For the first, CANX-7 is the chosen mission, as it was a successful one with publicly available data. CANX-7’s main objective was to demonstrate the use of an external drag sail module. It was launched in a Sun Synchronous Orbit (SSO) with an altitude of 688 km and an inclination of 98°. The mission utilized an ADS-B receiver as its payload. The 3U (10 cm x 10 cm x 34 cm) CubeSat platform weighed 3.75 kg and relied on a purely magnetic attitude determination and control system. Of the 3U volume, 2U was allocated to spacecraft housekeeping and payload systems, while 1U housed its four drag sail modules stacked in pairs. The objective of the drag sail module was to facilitate the quickest possible deorbiting, in alignment with the Inter-Agency Space Debris Coordination Committee (IADC) guidelines, which recommend a maximum post-mission deorbit time of 25 years \cite{NASA2020}. In practice, the expected deorbit time for CANX-7 was much shorter, estimated at around 5 years following the deployment of the drag sail in 2017.

Considering these characteristics of the CANX-7 mission, the primary objective of the ADM in CubeSat configuration is to deorbit a CubeSat up to 10U in size from lower earth orbits up to 2000 km. Additionally, it aims to contribute to the development of COTS PnP passive deorbiting systems for CubeSats and the advancement of drag sail-induced deorbiting technology. A size constraint of 1U (10 cm x 10 cm x 10 cm), implying a mass constraint of 1 kg, is set for this configuration. The design must also comply with CubeSat design specifications and secondary payload launching systems. The use of space-tested components and materials is mandated to ensure easy development, reduce costs, and minimize development time. Other functional requirements due to the PnP nature of the module include i. The ADM module should be easily attachable, with adaptable dimensions for host satellites. ii. It should be independently deployable and act as a completely independent system, utilizing minimal host satellite power and components.

While CANX-7 serves as a reference mission for the CubeSat configuration of ADM, the design is not limited to missions of the CANX-7 type. This configuration is intended to demonstrate adaptability to other spacecraft platforms, such as the Generic Nanosatellite Bus (GNB) and the Nanosatellite for Earth Monitoring and Observation (NEMO) bus designs developed by the University of Toronto Institute for Aerospace Studies Space Flight Laboratory (UTIAS SFL). An important requirement is the ability to test the design in a 1g environment or another easily reproducible controlled environment \cite{Pranajaya2010}.

The second configuration targets satellites larger than 10U (1m x 1m x 1m) in volume and greater than 10 kg in mass. Although numerous missions have successfully used drag sails for deorbiting smaller satellites and upper stages, the use of drag sails for deorbiting larger satellites (e.g., over 500 kg) is still an emerging area. For example, missions like dragNET have demonstrated the viability of drag sails for larger objects, though comprehensive data on deorbiting very large satellites using drag sails remains limited. Thus, the results from the study ‘ADEO De-Risk Dynamic Analysis’ (ADDA) \cite{Sinn2018} conducted by ESA are used for reference. The ADDA aimed to verify the feasibility of deorbiting a spacecraft using a drag sail. One of its findings was that a 25 m\(^2\) sail on a 300 kg satellite from a 600 km orbit could reduce the satellite's post-mission lifetime by 97\%, deorbiting in 5 years instead of 140 years.

The final requirements are derived from a combination of the individual necessities of both configurations. The objective is to deorbit a satellite with a mass of up to 200 kg from an orbit of up to 2000 km altitude or an equivalent ballistic coefficient. A size constraint of 2U (20 cm x 10 cm x 10 cm), with a mass constraint of 2 kg, is set. The design is ensured to be composed of COTS components, which are space-tested, to facilitate development and minimize costs and time. Like the CubeSat configuration, this configuration must also be easily attachable, adaptable in dimensions to host satellites, independently deployable, and act as a completely independent system with minimal power requirements.

Both configurations must withstand the harsh environmental conditions in space, including Atomic Oxygen (ATOX), ultraviolet (UV) radiation, and temperature extremes.

\begin{figure}[ht!]
  \centering
  \includegraphics[width=0.5\textwidth]{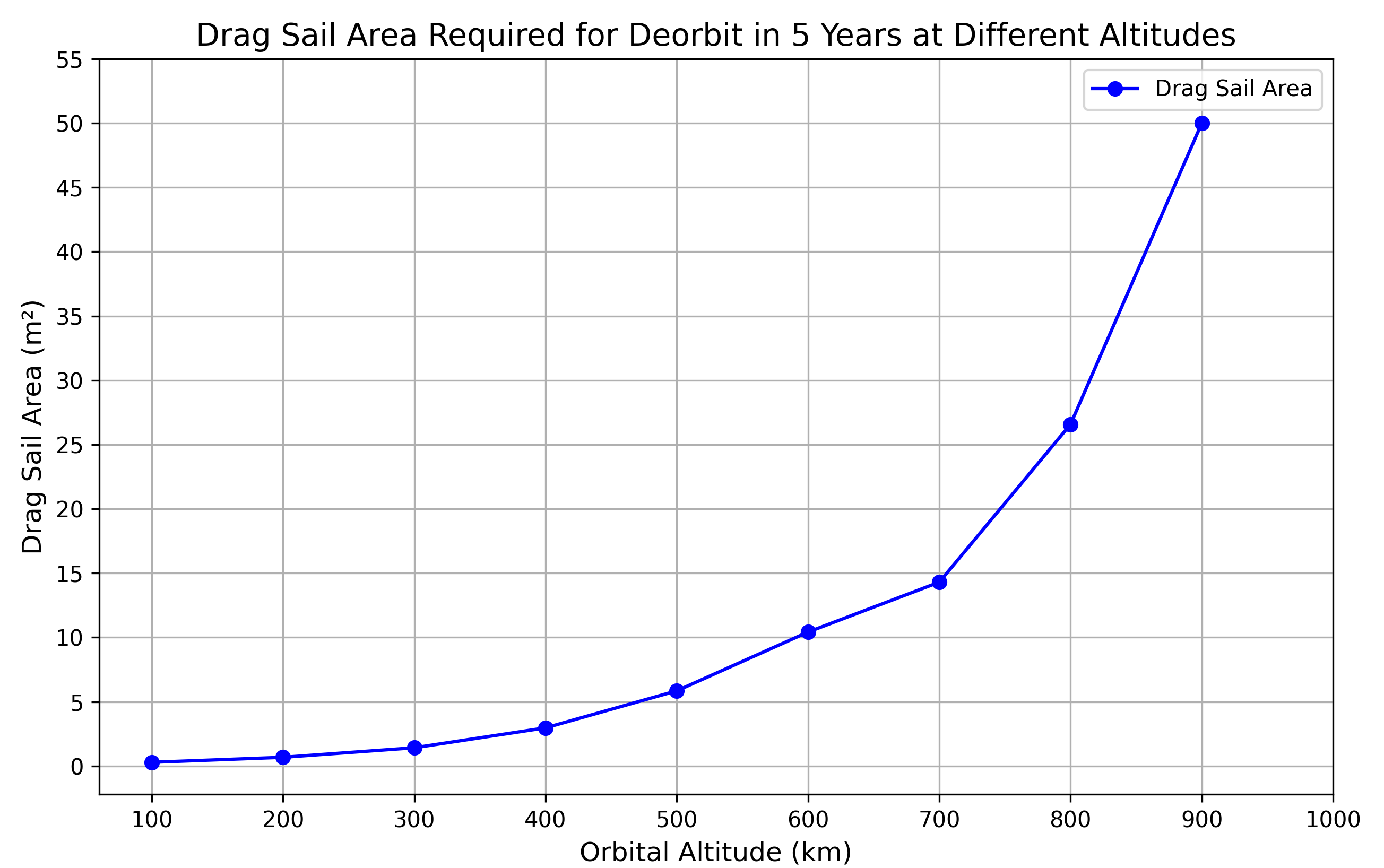}
  \caption{Drag Sail Area vs Altitude for deorbit in 5 years}
  \label{fig:dsa_5year}
  \vspace{-1em} 
\end{figure}

In accordance with the new FCC guidelines, the required drag sail area for deorbiting within 5 years is shown in Figure \ref{fig:dsa_5year}. 

\begin{figure}[H]  
  \centering
  \includegraphics[width=0.5\textwidth]{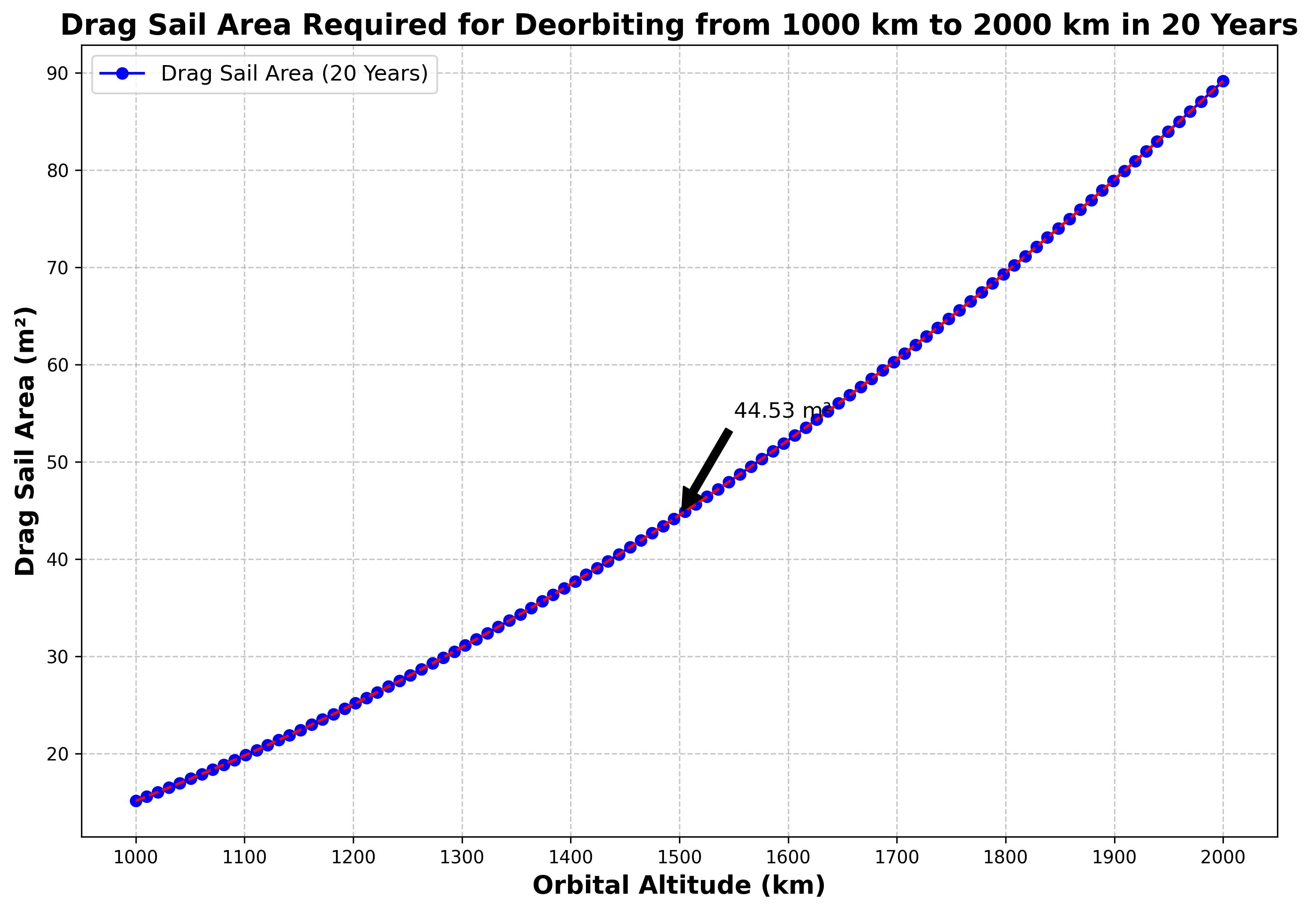} 
  \caption{Drag Sail Area vs Altitude for deorbit in 20 years}
  \label{fig:dsa_20year}
\end{figure}

However, beyond 900 km, the required drag sail size becomes impractically large for deorbiting within 5 years. Therefore, Figure \ref{fig:dsa_20year} presents the required drag sail area for deorbiting within 20 years for altitudes between 1000 and 2000 km. This approach ensures timely deorbiting while avoiding excessively large cross-sectional areas that could pose a collision risk to other satellites in similar orbits. \par

To better communicate the design requirements, we use the concept of the ballistic coefficient (BC), which effectively integrates both mass and drag area into a single metric. This provides a more comprehensive understanding of the satellite's deorbiting performance compared to using mass and drag areas separately.

The ballistic coefficient is defined as:

\begin{equation}
BC = \frac{m}{A C_d}
\end{equation}

where \( m \) is the mass of the satellite, \( A \) is the cross-sectional area perpendicular to the direction of motion, and \( C_d \) is the drag coefficient. For low Earth orbit (LEO) satellites, \( C_d \) is typically around 2.2.

Additional equations for context include:

\begin{equation}
a = \frac{F_d}{m} = \frac{1}{2} \frac{\rho A C_d v^2}{m}
\end{equation}
where \(a\) is the acceleration due to drag (in m/s\(^2\)).

\begin{equation}
t = \frac{2 m}{\rho A C_d v}
\end{equation}

where \(t\) is the orbital decay time (in s).

These equations help in assessing the effectiveness of the drag sail in achieving the desired deorbiting performance within the constraints of safe operational practices.

\section{Optional Increase In Deorbit Rate}

\begin{figure}[h]
    \centering
    \includegraphics[width=\linewidth]{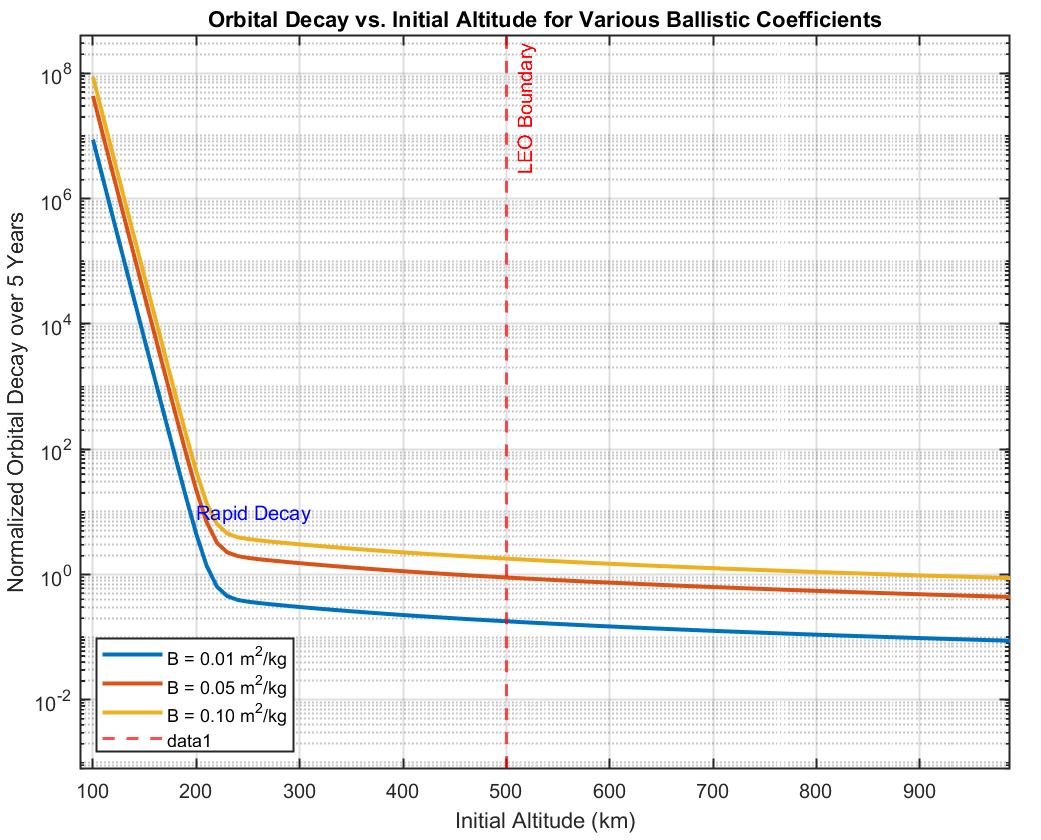} 
   
    \caption{Orbital Decay vs. Initial Altitude for different BCs}
    \label{fig:orbitaldecay}
\end{figure}

Figure \ref{fig:orbitaldecay} illustrates how orbital decay varies with initial altitude for different ballistic coefficients, showing that lower altitudes and higher ballistic coefficients result in more rapid decay. 

\par
To increase the rate of deorbit by ADM, simulations were performed. The results are combinations of fixed-attitude and variable-attitude simulations. The varying attitude simulation results are needed as the fixed-attitude simulations fail to account for the time-varying nature of the spacecraft's projected drag area. This could be accounted for using an active control system, which is not viable for the lifetime of deorbit. Aerodynamic perturbations are not accounted for below 450 km, nor is geomagnetic disturbance due to residual magnetic dipole moments above 650 km. Assuming a constant drag area could yield very misleading results, and thus fixed-attitude simulation results are needed.

\begin{figure}[h]
    \centering
    \includegraphics[width=\linewidth]{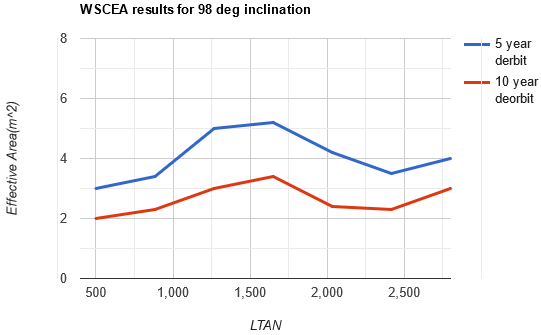} 
    \caption{WSCEA results for mentioned orbits and inclination}
    \label{fig:wscea}
\end{figure}

To assess this, short-term attitude simulations were used to evaluate the deorbit performance in slices of an entire $\sim$11-year solar cycle and combined appropriately to arrive at a result that approximates the performances. The output (WSCEA \cite{Shmuel2012}) can be compared against the constant area simulations performed on FreeFlyer\texttrademark{}. The results in figure \ref{fig:wscea} were obtained for two configurations: a 0-degree angle between the drag sailplane and residual dipole for a 5-year deorbit profile and a 90-degree angle for a 10-year deorbit profile. The orbit considered was at 700 km altitude with a 98.1-degree inclination and LTAN ranging from 0500 HHMM to 2800 HHMM.

These results indicate that providing occasional stabilization in the form of nadir pointing/PID control to keep the drag area constant would increase the rate of deorbit and thus decrease deorbiting time. The most effective stabilization occurs at $\sim$1000 HHMM LTAN and $\sim$2000 HHMM LTAN, as per the results, thus maximizing the duration of higher drag area. Similarly, stabilizations could be performed following simulations of respective missions using the ADM in the future.

\section{Key Differences From Existing Technologies}
Several key distinctions set our invention apart from existing drag sail deployment and deorbiting technologies.

One approach described in US20200130872A1 \cite{US20200130872A1} incorporates hubs that stow triangular sections of a drag sail. Each hub, equipped with retractable booms and stepper motors, extends its section of the sail upon deployment in space. In contrast, our design uses a rotator—a custom-built stepper motor—around which the sail is wrapped. Deployment is triggered by achieving a specific angular momentum, calculated based on the satellite’s specifications. Unlike the segmented deployment method in US20200130872A1, our sail is deployed as a single, continuous piece, offering a more streamlined solution.

In another method detailed in WO2009149713A1 \cite{WO2009149713A1}, the focus is on controlling the shape of flexible elements using electromagnets. This technique, while applicable to drag sail deployment, primarily addresses shape control rather than providing a complete deorbiting solution. Our invention, on the other hand, integrates the sail deployment into a comprehensive deorbiting system, distinguishing it from techniques that only control the shape of the sail.

Additionally, WO2009149713A1 describes a method similar to the mechanical energy storage techniques found in some designs. These methods involve wound-up structures to deploy a segmented sail. Our invention diverges by utilizing a continuous sail deployment mechanism, thus offering a more efficient and integrated solution compared to the segmented approaches described.

These differences underscore the novel aspects of our design, emphasizing improvements in deployment efficiency and overall deorbiting system integration.
\section{AirDragMod (ADM) Design Overview}

This section presents the design of the AirDragMod (ADM), carefully thought through several iterations. The main stages can be distinguished: the conceptual design, the deployment mechanism, and the design overview. These phases are briefly introduced in the following subsections.

\subsection{Conceptual Design}

During the conceptualization of ADM’s design, the selection of key configuration elements is guided by figures of merit derived from mission analysis. Three main elements studied are the drag sail, the rotator, and the ADM structure layout. The design criteria followed during the decision-making process have been influenced by six key aspects: 1. fulfillment of deorbit requirements, 2. better approach to deployment than existing methods, 3. maximum drag sail area while minimizing ADM volume occupied, 4. generation of a PnP module, 5. preference for COTS components and space-tested technology, 6. manufacturable design at minimum cost.

The ADM has been designed in accordance with the results. ADM solves the problems associated with current deployment techniques, specifically the stored mechanical energy boom deployment method used in CANX-7. The ADM is based upon the Solar sail model used in the IKAROS mission by JAXA. The idea behind the deployment was that the angular momentum of the module would cause four mass blocks connected with sail ends to extend due to centrifugal force.

Both configurations of the ADM use a similar technique to deploy their stowed sails. Based on this and the previously obtained results, four critical components were chosen: 1. Sail, 2. Rotator, 3. Connector.

\begin{figure}[h]
\centering
\includegraphics[scale=0.5]{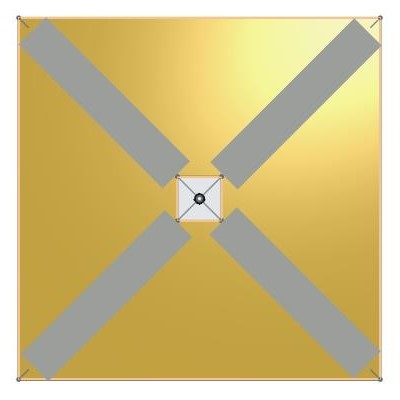}
\caption{Photopolymer application at diagonals of sail}
\label{fig:photopolymer_application}
\end{figure}

\begin{figure}[h]
\centering
\includegraphics[scale=0.5]{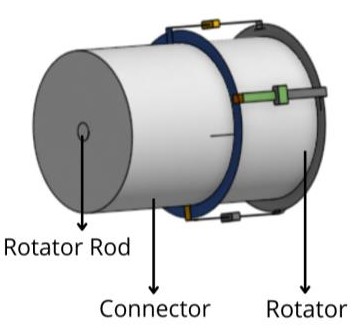}
\caption{Connector, Rotator, Rotator Rod}
\label{fig:connector_rotator_rotor}
\end{figure}

 The first element chosen encompasses the sail configuration and the characteristics of its membrane. The influence of this element of the design requires the selection of the configuration of the sail. A square-shaped configuration was chosen due to providing the best drag to mass ratio as well as having a broad-space heritage. The chosen sail configuration is a single unit with intrinsic divisions into four triangular quadrants. This configuration is perfect for deployment using angular momentum as the individual mass tips could uniformly spread out the sail and stay at a maximum distance on its diagonal. The average size of the sail to be used would stay 7m\(^2\). The proposed material of the sail is a metalized polymer similar to the one used for the sail of the CANX-7 mission. The polymer proposed to be used is a 12.7-micrometer Kapton\texttrademark{} film with a 300-Angstrom Aluminum coating capable of handling greater than 200° temperatures. This makes the mass/m\(^2\) equal to $\sim$7 for CubeSat configuration and $\sim$4 for satellite configuration. The aluminum coating allows for achieving high equipotential spacecraft structures and avoids excessive charging of the membrane. The material selection was in accordance with the COTS philosophy and the objective of using tested components. Secondly, since no boom deployment mechanism will be used, the absence of boom configuration would otherwise cause the deployed sail to fold in on itself as the masses lose tension. The spacecraft will not be kept in a constant revolving state as the increase in mass would mean increased power usage to keep a controlled spin rate; it increases communication hindrances. To solve this, a photopolymer in the form of epoxies or nitrile rubber would be used in the sail along its diagonals (fig. \ref{fig:photopolymer_application}). Photopolymers harden under sunlight, undergoing a process called curing. The material selection again follows the COTS philosophy, and a preference for those already tested in space would be given. 
The second element of importance is the Rotator. It derives aspects of stowage from the IKAROS probe design. The primary functions of the Rotator include stowage and deployment. The Rotator element is a cylindrical probe of diameter and length not exceeding 20 cm in both configurations. The sail is folded in a rolled 4 petal fashion similar to IKAROS. The IKAROS deployment mechanism utilizes the 4 stoppers, which aid in the deployment of the rolled petals using a relativistic rotation mechanism (motor drive) and eventually are released to allow the expansion of the fully deployed petals into a square sail. The tip masses attached to sail ends are also stowed in the Rotator and are the first ones to deploy on rotation, aiding in the deployment process by the tension created due to experiencing centrifugal force. Each tip mass is 50g in mass. Angular momentum to deploy the sail is generated by a reaction wheel. Adhering to COTS and space-tested component requirements, the reaction wheel suggested is CubeSpace’s CubeWheel\texttrademark{} Small+. It is the only rotating module and is further connected to the host spacecraft by a Connector module. 
The Connector module is a passive module, not present in the CubeSat configuration, connecting the Rotator to the host satellite. It is fixed to the satellite on one end and extends a cylindrical rod until the end of the Rotator. The Rotator rotates around this rod of connector as the axis. The Connecter furthermore houses a housekeeping computer controlling the entire module primarily providing needed power control. To dump excess momentum and avoid tumbling in the satellite caused by this rotation, the connector consists of magnetorquer rods. Components chosen based on COTS philosophy and space-tested systems are NCTR-M002 Magnetorquer Rod. This isn't present in the CubeSat configuration as the Rotator is directly attached to the satellite if it is capable of holding together during the deployment phase. This is done to keep the mass and power usage low. Figure \ref{fig:connector_rotator_rotor} shows the design of the Connector and Rotator.

\subsection{Petal Model}
The equation of motion (domain equation) for extended sail petal is \cite{Stutts2000}
\begin{equation}
\frac{\partial}{\partial x} \left( \tau(x) \frac{\partial u}{\partial x} \right) + f(x,t) = \rho(x) \frac{\partial^2 u}{\partial t^2}
\end{equation}

\begin{figure}[h]
\centering
\includegraphics[scale=0.3]{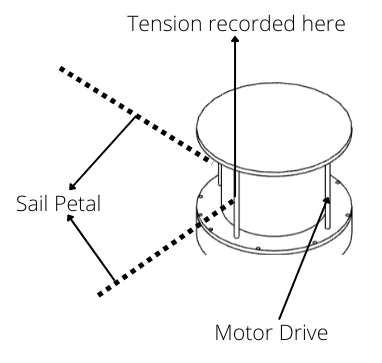}
\caption{Location of Flex Sensor (tension)}
\label{fig:location_of_flex_sensor}
\end{figure}

This equation, along with its boundary values and initial values, forms the initial value boundary problem and can be used to model the motion in Simulink\texttrademark{}. In the equation, \(\tau\) is a function of tension varying with position on the petal. The sail petal can be assumed to be a continuous mass distribution; the mass of the sail spreads over the length of the petal. Figure \ref{fig:location_of_flex_sensor} provides a representation of the location where tension data is recorded. Using Newton’s second law and definite integration over a small mass element, \(\tau\) can be found to be:
\begin{equation}
\tau(x) = \frac{M}{2L} \omega^2 (L^2 - x^2)
\end{equation}

\begin{figure}[h]
\centering
\includegraphics[width=0.9\linewidth]{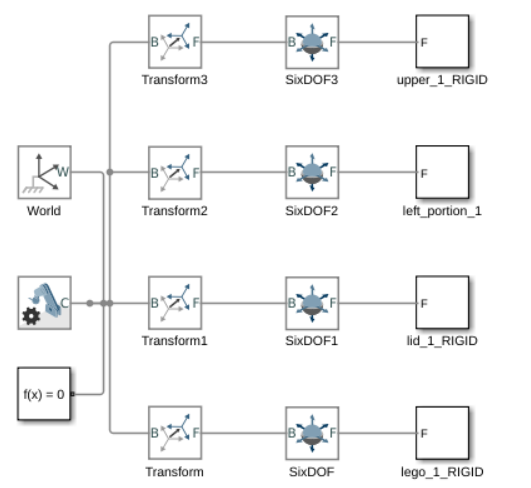}
\caption{SimScape Physical System Block Diagram}
\label{fig:simscape_physical_system}
\end{figure}

Here, \(\frac{M}{L}\) is the linear mass density; for a sail, it is \(7 \, \text{m}^2\). The linear mass density of the petal is \(6.2756 \, \text{g/m}\) if the material chosen is Aluminized Kapton\texttrademark{} Polymer \cite{dupont_kapton}. Here, the position \(x\) is a function of time, which depends on the rate at which motor drive units release the petals. This rate is dependent on the deployment time and the length of the sail petal to be extended. For a \(7 \, \text{m}^2\) sail, this rate during the 120-second sail release period (for the setup mentioned earlier) is \(4.124 \, \text{cm/s}\). \(\omega\) is the angular velocity of the rotator unit. The model of the physical system, i.e., the ADM module in Simulink\texttrademark{}, is shown in Figure \ref{fig:simscape_physical_system}. This model represents the physical system over which active control is to be achieved.


\subsection{Deployment Sequence}
\begin{figure}[h]
\centering
\includegraphics[width=\linewidth]{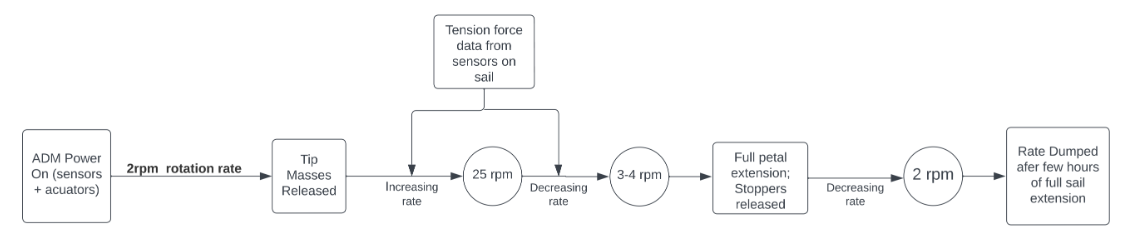}
\caption{Standard Deployment Sequence}
\label{fig:standard_deployment_seq}
\end{figure}

The deployment process can be initiated from the ground or host spacecraft. Once initiated, the power supply to the reaction wheel is regulated and increased to have the Rotator rotate at 5 rpm over a span of 6 seconds. As the rotation speed increases gradually, at 2 rpm, the tip masses are released, which are clutched mechanically. This creates a centrifugal force experienced by the tip masses which extends the sail-petals gradually connected to the Rotator by tethers. The spin rate is now increased to 25 rpm as the petals extend out while stoppers hold the membrane through the relative rotation mechanism. The spin rate gradually decreases as the petals extend half their size to $\sim$15 rpm as no new torque is being produced. The spin rate on full extension of petals is $\sim$3 to 4 rpm, and on full extensions, the stoppers are released. As the stoppers “fall down”, the final stage of deployment begins and the sail starts acquiring its square shape from petals. By this point, the rpm is low at $\sim$2. In CubeSat configuration, this rate is kept for an hour under sunlight post-deployment for the sail to harden at its diagonals, after which the dumping devices are used to dump the rotation. In the other configuration, the Rotator is kept spinning at the rate as the Connector’s dumping magnetorquer dumps excess rotation and draws power from the satellite. Figure \ref{fig:standard_deployment_seq} illustrates the standard deployment sequence as originally conceived, representing the general unmodified approach.

\subsection{Sample Cost Estimates}
To fulfill its initial objectives, the design should provide the said results within the cost and mass constraints, minimizing both of them. An estimation of cost and mass for both configurations is provided in Table 1, which is located at the end of this document. This table also justifies the division of the design into two configurations, as evidenced by the considerable mass and cost differences mentioned, apart from the size difference described in the Conceptual Design. The components are COTS and space-tested on multiple missions previously, thus adhering to the initially stated philosophies.
\begin{table}[h!]
\centering
\begin{tabular}{|l|c|c|}
\hline
\textbf{Component} & \textbf{Mass} & \textbf{Estimated Cost (USD)} \\ \hline
\multicolumn{3}{|c|}{\textbf{Cubesat Config}} \\ \hline
Chassis & 700g & 2500 \\ \cline{1-3}
Reaction wheel & 90g & 5950 \\ \cline{1-3}
Sail & $\sim$50g & 500 \\ \cline{1-3}
Dumping rod & 30g & 1200 \\ \cline{1-3}
Electronics & 1500g & 700 \\ \cline{1-3}
\textbf{NET} & \textbf{$\sim$1100g} & \textbf{$\sim$10k USD} \\ \hline
\multicolumn{3}{|c|}{\textbf{Sat Config}} \\ \hline
Chassis & 1000g & $\sim$3100 \\ \cline{1-3}
Reaction wheel & 90g & 5950 \\ \cline{1-3}
Sail & $\sim$28g & 250 \\ \cline{1-3}
Dumping rod & 30g & 1200 \\ \cline{1-3}
Electronics & 200g & 800 \\ \cline{1-3}
\textbf{NET} & \textbf{$\sim$1400g} & \textbf{$\sim$11k USD} \\ \hline
\end{tabular}
\caption{Component Mass and Estimated Cost for Cubesat and Sat Configurations}
\label{tab:component_costs}
\end{table}

\section{Results and Discussion}
\begin{figure}[h]
\centering
\includegraphics[scale=0.3]{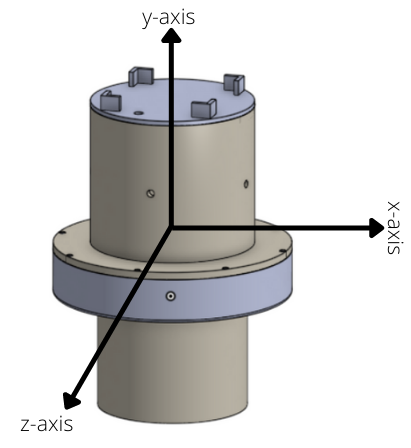}
\caption{Setup CAD model with local axes}
\label{fig:Setup_CAD_model_with_local_axes}
\end{figure}

Owing to hardware limitations, the force models do not account for lateral forces on sail petals during deployment, nor are the disturbance models computed in the y-axis of the system as the setup was placed on a surface and not in a simulated microgravity environment. The force and acceleration profiles are two-dimensional but consistent with the expected three-dimensional model. Hence, the current modeling can be extended into three dimensions. 
The general setup, along with axes, is shown in the figure. Sail placeholders were used in the setup. Testing using actual sail material would yield the lateral force profile. The configuration geometry is novel to the design, however, the number of deployment motor drives can be varied with a minimum of 4. The setup consisted of 4 such motor drives. The release rate was kept as mentioned earlier. The tension data recorded at the prototype was at the base of motor drive units during the sail petal release period of the deployment sequence. Theoretically, tension would increase initially, then decrease to attain a constant value. The axes for the setup are shown in Figure \ref{fig:Setup_CAD_model_with_local_axes}. The roll axis for the setup is the y-axis; angular velocity is measured around it. Ideally, there should not be any movement in other axes, however, perturbations would be caused and thus need for an active control system. 
The curve-fit of tension force data was done in MATLAB\textsuperscript{TM}. The best-fit model equation obtained is given below, with the coefficient values given in Fig. \ref{fig:curve_fit_coefficient_values}
\begin{equation}
f(x) = a_1 \cdot \sin(b_1 x + c_1) + a_2 \cdot \sin(b_2 x + c_2)
\end{equation}
\vspace{-10pt}  
\begin{figure}[h]
\centering
\includegraphics[width=0.8\linewidth]{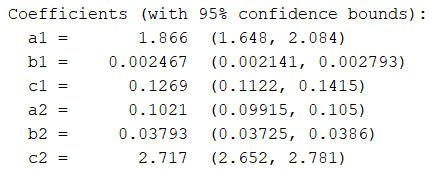}
\caption{Curve-fit coefficient values}
\label{fig:curve_fit_coefficient_values}
\end{figure}
\par
\begin{figure}[h]
\centering
\includegraphics[width=\linewidth]{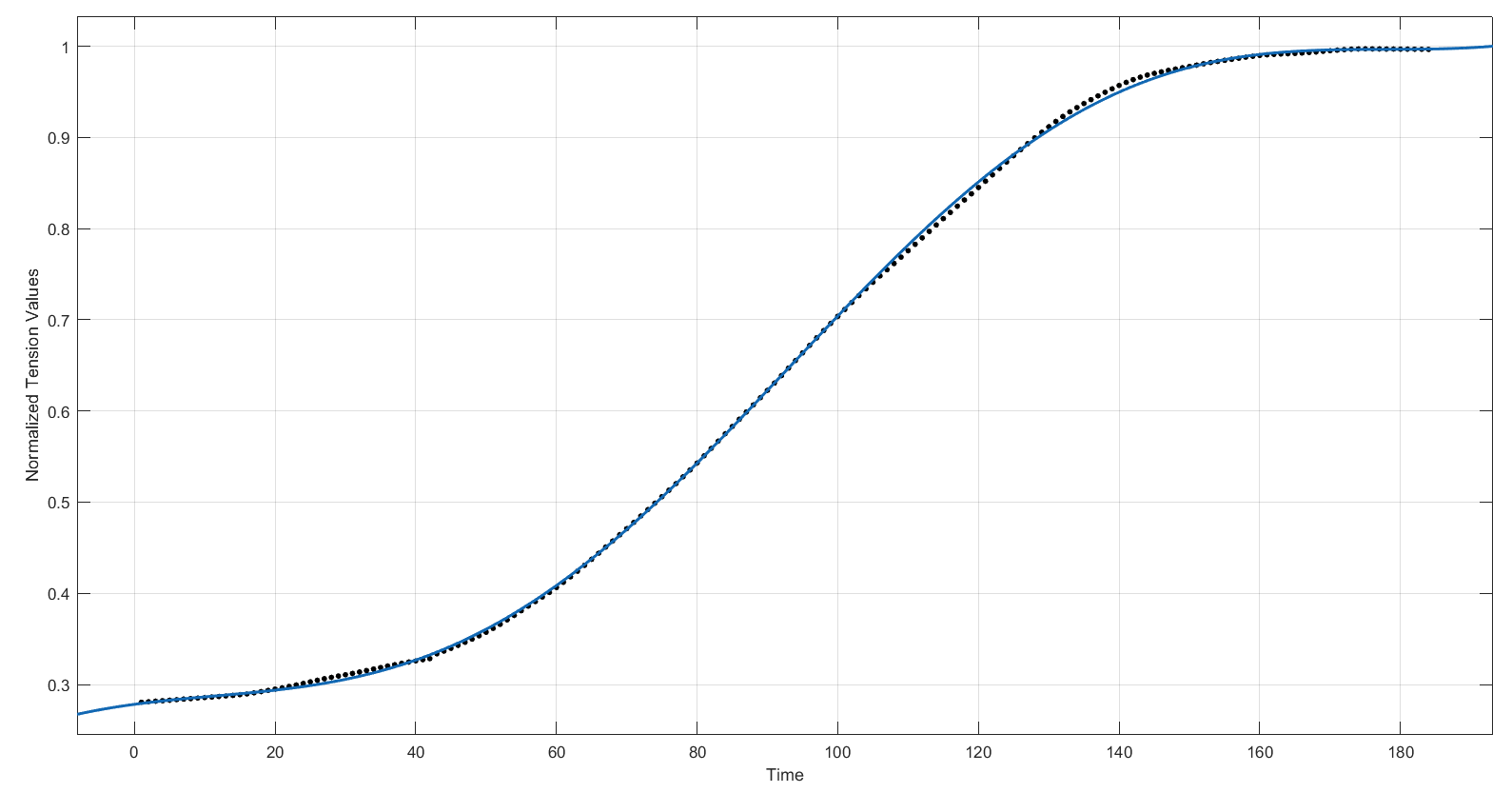}
\caption{Curve-fit Normalized Tension Values}
\label{fig:curve_fit_tension_values}
\end{figure}
\begin{figure}[h]
\centering
\includegraphics[width=\linewidth]{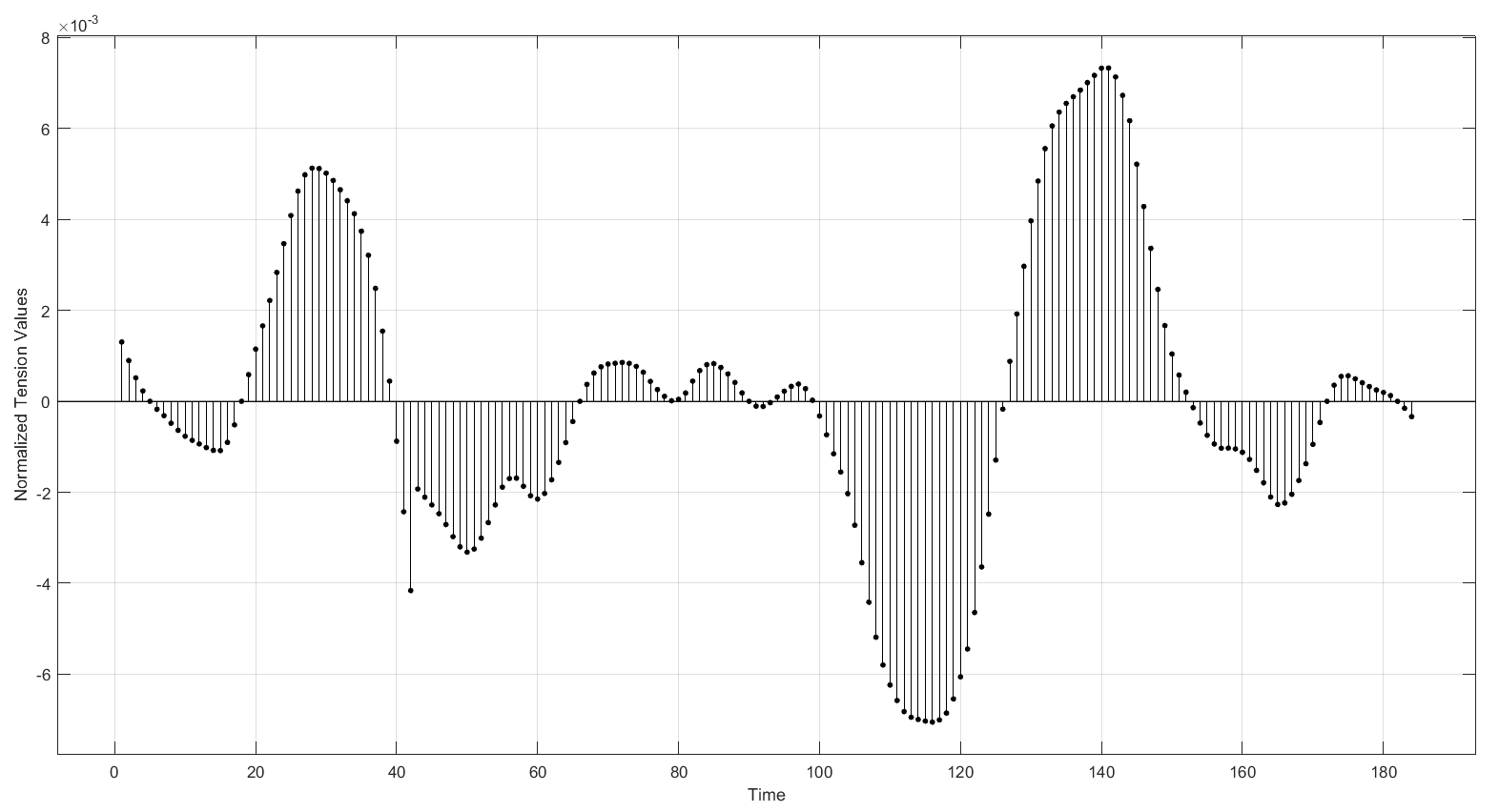}
\caption{Residuals Plot of figure \ref{fig:curve_fit_tension_values}}
\label{fig:residuals_tension_plot}
\end{figure}

Figure \ref{fig:curve_fit_tension_values} shows the curve-fit plot along with the residuals plot for the given fit is shown in figure \ref{fig:residuals_tension_plot}.
The Tension values are normalized moving means of values recorded over multiple test runs. 
\par
\begin{figure}[h]
\centering
\includegraphics[width=\linewidth]{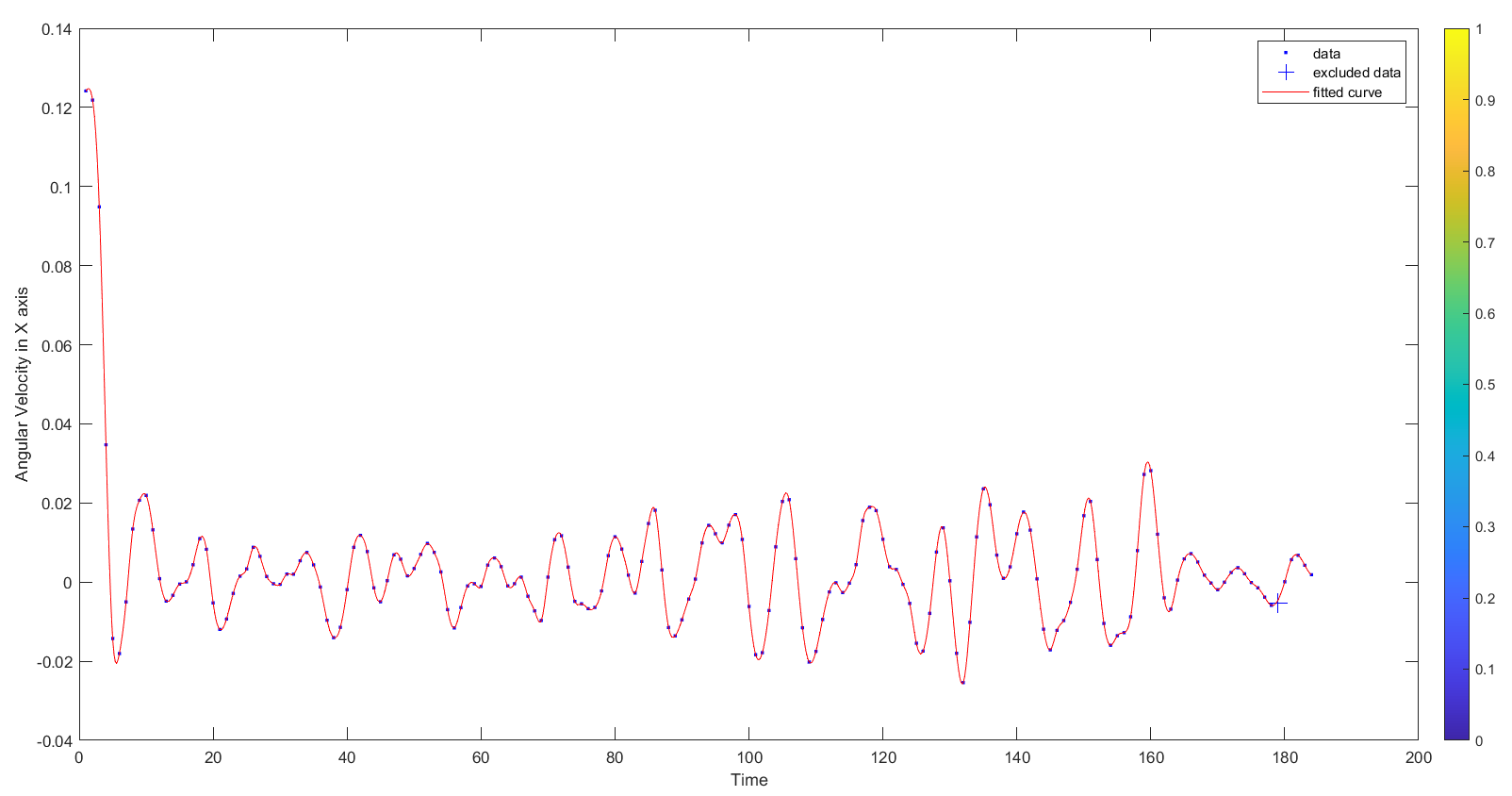}
\caption{Interpolant of Angular Velocity in X-axis vs. Time}
\label{fig:interpolant_angular_vel}
\end{figure}

The interpolation of angular velocity recorded using IMU in the x-axis is displayed in figure \ref{fig:interpolant_angular_vel}. The cubic spline interpolant in $x$ is a piecewise polynomial over p, where $x$ is normalized by mean 92.03, standard deviation 53.02, and p is a coefficient structure. From figure \ref{fig:interpolant_angular_vel}, it can be inferred that some periodic disturbance of oscillation nature is caused in the x-axis. This disturbance needs to be controlled by the control system under development. \par
\par
\begin{figure}[h]
\centering
\includegraphics[width=\linewidth]{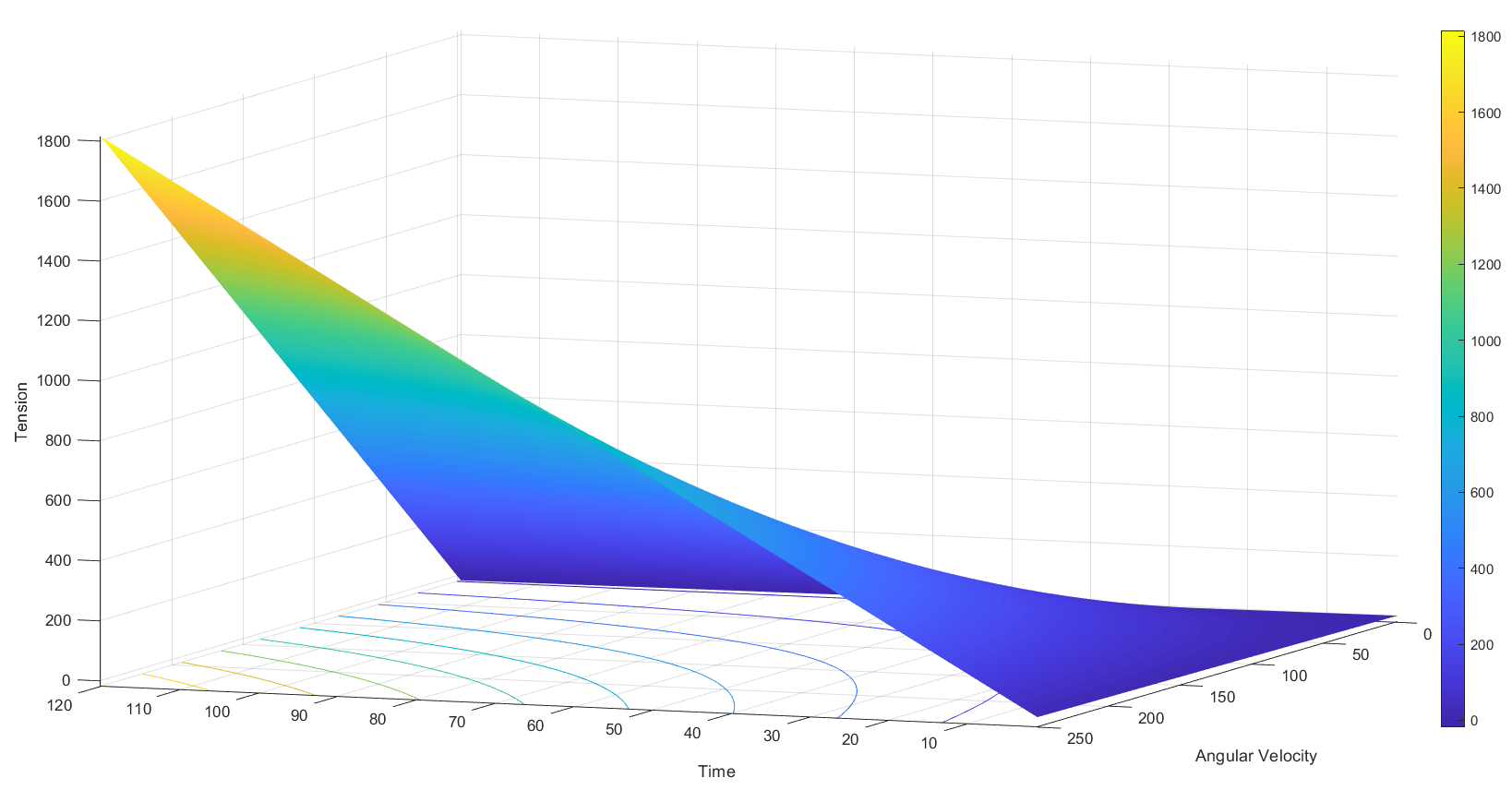}
\caption{Surface Plot with Time as X-axis, Angular Velocity as Y-axis, and Tension as Z-axis}
\label{fig:surface_plot}
\end{figure}

Figure \ref{fig:surface_plot} shows a surface and contour plot of time, angular velocity, and tension values obtained from the numerical model. 
\par
Incorporating the analysis of all figures provides a detailed assessment of the AirDragMod (ADM) system's performance during deployment:

Figure \ref{fig:curve_fit_tension_values} demonstrates the normalized tension values over time, starting with an initial sharp increase followed by a gradual rise, peaking around 120 seconds. This trend aligns with the expected deployment sequence, reflecting the behavior of the sail petals as they extend under increasing centrifugal forces.

Figure \ref{fig:interpolant_angular_vel}, which interpolates the angular velocity along the x-axis, reveals periodic oscillations. While these oscillations are small, they emphasize the need for an active control system to stabilize the deployment process.

Figure \ref{fig:surface_plot} presents a surface plot that correlates time, angular velocity, and tension, providing insight into the dynamic interplay during deployment. Higher tensions generally coincide with increased angular velocities and later deployment phases, reinforcing our design principles.

The residuals plot in Figure \ref{fig:residuals_tension_plot} reveals key information about the accuracy of our curve-fitting model. Although the residuals are of a small magnitude (in the order of \( 10^{-3} \), their non-random oscillatory pattern points to systematic deviations from the fitted curve. These oscillations are consistent with the periodic disturbances seen in the angular velocity data, suggesting a link between unaccounted-for tension variations and perturbations in rotational dynamics.

Notable deviations in the residuals occur around the 30-second, 110-second, and 140-second marks, which coincide with critical phases of the deployment sequence. This indicates the inherent complexity of the process and underscores the challenges in developing a fully comprehensive analytical model.

These combined results validate several aspects of the ADM design, while also highlighting areas that require further refinement, particularly in controlling perturbations and improving tension modeling. The data suggests that a more advanced model, incorporating higher-order terms or additional physical phenomena, may be necessary to capture the subtle yet significant variations observed during deployment.

\section{Conclusion and Future Work}
Going forward, the findings will guide the enhancement of our control algorithms, especially in adapting to unexpected perturbations during critical phases of deployment. Further investigation into the underlying causes of these residual patterns could lead to structural or mechanical improvements in the ADM system, potentially reducing these variations at their source.

The analysis not only supports our initial design concepts but also provides critical insights for future iterations. Addressing the root causes of these residual variations will enhance the accuracy of our models, leading to improved predictability and reliability of the ADM system across various orbital scenarios. This deeper understanding of system behavior will be pivotal in refining the active control system to ensure stable and consistent drag sail deployment across a wide range of mission profiles.

A detailed model for both designs is essential and will need to be carefully developed. Ultimately, a full prototype will be constructed, followed by exhaustive testing of deployment and the optional attitude control technique for maximizing drag.

\section{Glossary}

\begin{itemize}
    \setlength{\itemsep}{0pt} 
    \item LEO - Low Earth Orbits
    \item PnP - Plug and Play
    \item COTS - Commercial-Off-The-Shelf
    \item ADM - AirDragMod
    \item OCC - Orbital Control Capacity
    \item IADC - Inter-Agency Space Debris Coordination Committee
    \item LTAN - Local Time of Ascending Node
    \item UTIAS SFL - University of Toronto Institute for Aerospace Studies Space Flight Laboratory
    \item ATOX - Atomic Oxygen
    \item UV – Ultraviolet
    \item WSCEA - Whole Solar Cycle Effective Area
    \item ESA - European Space Agency
    \item LEO\(_{\text{IADC}}\) - IADC LEO protected region with \( h \in [0, 2000] \), where \( h \) refers to altitude
\end{itemize}

\bibliographystyle{unsrt}
\bibliography{references}

\end{document}